# MoS$_2$-on-paper optoelectronics: drawing photodetectors with van der Waals semiconductors beyond graphite

Ali Mazaheri[a,b], Martin Lee[c], Herre S.J. van der Zant[c], Riccardo Frisenda[a], Andres Castellanos-Gomez[a,*]

We fabricate paper-supported semiconducting devices by rubbing a layered molybdenum disulfide (MoS$_2$) crystal onto a piece of paper, similarly to the action of drawing/writing with a pencil on paper. We show that the abrasion between the MoS$_2$ crystal and the paper substrate efficiently exfoliates the crystals, breaking the weak van der Waals interlayer bonds and leading to the deposition of a film of interconnected MoS$_2$ platelets. Employing this simple method, that can be easily extended to other 2D materials, we fabricate MoS$_2$-on-paper broadband photodectectors with spectral sensitivity from the ultraviolet (UV) to the near-infrared (NIR). We also used these paper-based photodetectors to acquire pictures of objects by mounting the photodetectors in a homebuilt single-pixel camera setup.

## Introduction

Handwriting and/or drawing on a piece of paper with a pencil has become a routine daily task for thousands of millions of people around the world due to their mass production that led to ubiquity and reduced cost. These common stationary items have recently jumped out of the writing/drawing realm and have been employed to fabricate electronic devices. This has been, most likely, motivated by the extremely low cost of paper substrates (paper ~0.1 €/m$^2$ as compared with PET ~2 €/m$^2$, PI ~30 €/m$^2$, crystalline silicon ~1000 €/m$^2$),[1,2] its biodegradability and its potential to allow the fabrication of flexible and even foldable electronic devices.[1,3–5]

The rough, fiber-based, structure of paper, however, is a handicap to fabricate devices using conventional lithographic techniques developed to fabricate devices on silicon wafers by the semiconductor industry. The use of graphite pencil lead traces, formed by the exfoliation of graphite platelets through the abrasion of the graphite lead while scribing on the paper substrate, allow to pattern electrically conductive pads on the rough surface of paper.[3,6,7] This simple approach has been used to demonstrate pencil-drawn-on-paper strain gauges, humidity, temperature, gas and chemical sensors.[6,8–15]

The lack of band gap in graphite, however, hampers the use of pencils to fabricate certain electronic devices, like digital electronics components or optoelectronics devices that require a semiconductor material with a sizeable band gap. Nonetheless, the amount of works studying draw-on-paper semiconductor devices is still very scarce.[16–20]

Here, we explore the potential of layered semiconducting materials to draw semiconductor devices through abrasion on paper. We select molybdenum disulfide (MoS$_2$) as an illustrative example of a van der Waals semiconductor, which is abundant in nature in the mineral form of molybdenite. We demonstrate that the layered structure of MoS$_2$, very similar to that of graphite, allows for drawing MoS$_2$ platelets traces on paper by simply rubbing a MoS$_2$ crystal against a paper substrate. We show the potential of the fabricated devices as broadband photodetectors with spectral sensitivity in the ultraviolet (UV), visible (VIS) and near-infrared (NIR) range. Moreover, the paper-based photodetectors can be used to acquire pictures of objects by integrating these photodetectors in a single-pixel camera setup. The fabrication process described here is a general one and could be straightforwardly applied to other van der Waals semiconductors opening a wide field of research.

## Results and discussion

### Paper-supported device fabrication

The device fabrication starts by printing the electrodes outline with a laser printer (Brother MFC-L5700DN) on conventional Xerox paper (80 g/m$^2$), see Figure 1.1. The semiconductor channel area is delimited by using scotch tape (3M, MagicTape®) to make a square mask. A freshly cleaved molybdenite crystal (Wolfram Camp mine, Queensland, Australia) is rubbed against the unmasked paper to form a homogeneous coverage, see Figure 1.2. In order to improve the homogeneity of the MoS$_2$ film we blur the as-drawn MoS$_2$ film with a cotton swab, Figure 1.3. We repeat the MoS$_2$ rub + blur steps 4 times until obtaining a highly homogeneous MoS$_2$ film, see Figure 1.4. We then remove the tape mask and we fabricate graphite electrodes by drawing with a 4B pencil (with an approximate composition of ~80% graphite, ~15% clay and ~5% wax)[21] filling the printed outline, Figure 1.5. We address the readers to Refs. [6,21] for a comprehensive characterization of the morphology, the chemical composition and the electrical properties of pencil-drawn electrodes. Interestingly, Figure 1.5 shows that it is possible to 'draw' different van der Waals materials on top of each other to build multi-layered structures with dissimilar 2D materials. This can be of interest for the fabrication of vertical devices or electronic components. Figure 1.6 shows an optical picture of a finished device. In order to solder wires, two pieces of electrically conductive copper tape are adhered on top of the graphite pads.

We find that the contact between the graphite electrodes and the $MoS_2$ channel is Ohmic with a contact resistance of ~20 MΩ (see the Supporting Information for details about the contact resistance measurement through a transfer length method). We also determine the sheet resistance of our devices which ranges from 0.5-10 GΩ/sq. Considering an average $MoS_2$ film thickness of 15-25 μm (see Figure 2 in the next section) we estimate that the conductivity of the drawn $MoS_2$ on paper films is in the $3\cdot10^{-6}-1.5\cdot10^{-4}$ S/m range. Interestingly, this conductivity range is higher than that reported for networks of liquid phase exfoliated $MoS_2$ ($6\cdot10^{-7}$ - $5\cdot10^{-5}$ S/m),[22–24] most likely due to the presence of solvent residues at the interfaces of liquid phase exfoliated materials that impair electrical conductivity.

**Morphological and compositional characterization**

We have characterized the as-drawn $MoS_2$-on-paper by scanning electron microscopy (SEM) and energy dispersive X-ray spectroscopy (EDX) in a FEI Helios G4 CX system. Copper tape was used to secure and electrically ground the $MoS_2$-on-paper sample during the measurement. The sample was tilted to the maximum angle of 52º to characterize the cross-section of the sample. An electron energy of 5 keV was used for imaging and EDX spectroscopy.

Figure 2a-c show SEM images with increasing magnification of the $MoS_2$ film on paper, far from the sample edge. We found that the $MoS_2$ is evenly deposited on the surface of the paper (as bare, uncoated, paper would show up with a very different contrast in the SEM image), coating the paper fibers and filling in the gaps between the fibers. On top of the fibers it is not possible (apart from some loose flakes) to resolve individual flakes nor layered structures, probably because the $MoS_2$ abrasion is so severe that the resulting film is composed of small crystallites (Figures 2a and 2b). Inside the gaps between the paper fibers, however, it is possible to resolve $MoS_2$ flakes with lateral sizes in the 0.5-4 μm range, displaying terraced structures (Figure 2c). By imaging at the edge of the paper substrate an estimate of the average film thickness can be obtained, which is in the 15-25 μm range (see Figures 2d and 2e). We have also carried out EDX spectroscopy for an insight into the composition of the as-drawn $MoS_2$ film (Figure 2f). The EDX spectrum of the $MoS_2$ film on paper presents a prominent peak at 2.3 keV, that corresponds to the Mo and S elements, and also shows two strong peaks associated with the presence of C and O. We attribute the presence of these peaks to the underlying paper substrate (see the EDX spectrum acquired on the uncoated paper for comparison as displayed in the top panel).

**Optical characterization**

We have characterized the optical properties of the $MoS_2$-on-paper films by Raman and transmittance spectroscopy. Raman measurements were carried out with a Renishaw InVia Reflex system with a 515 nm diode laser excitation, a magnification of 20×, a spot size of ~3 μm, a laser power of 5 mW and a notch filter at the laser line. Figure 3a shows the Raman spectrum acquired on the $MoS_2$-on-paper film. The figure also displays the Raman spectra acquired on a $MoS_2$ bulk crystal (purchased from SPI supplies®) for comparison. The $MoS_2$-on-paper film displays two prominent (and sharp) peaks corresponding to the $A_{1g}$ and $E^1_{2g}$ Raman modes (Figure 3a), indicating a high crystallinity of the as-deposited $MoS_2$. The Raman shift difference between both peaks implies that the film is composed by $MoS_2$ flakes >4 layers thick.[25–27] Interestingly both peaks are shifted towards lower values with respect to the bulk $MoS_2$ Raman modes. A similar simultaneous shift of these two Raman modes has been observed in turbostratic $MoS_2$ layers, suggesting that the abrasion forces during the rubbing of the $MoS_2$ crystal against paper breaks the interlayer van der Waals bonds leading to the slippage of atomic planes and forming a film with $MoS_2$ flakes with faulty stacking ordering.[28]

The optical transmission spectroscopy study has been performed using a high-intensity fiber coupled halogen illuminator (OSL2, Thorlabs) and a thermoelectrically cooled CCD spectrometer (#86-406, Edmund Optics). The spectrum of bare (uncoated) paper and the $MoS_2$ covered paper are acquired at identical illumination power and acquisition settings (integration time, averaging, etc). Figure 3c shows the transmission spectrum (plotted as 1-transmittance) where one can see three peaks, corresponding to the A, B and C excitonic resonances, and beyond the A exciton a long decay tail for longer wavelengths, indicating indirect-gap absorption. The position of the A, B and C exciton peaks (669 nm, 609 nm and 485 nm respectively) indicates that the nanosheets forming the film are composed of 3-6 layers thick $MoS_2$ flakes turbostratically stacked.[29,30]

**Optoelectronic characterization**

The performance of the fabricated $MoS_2$-on-paper devices as photodetectors is studied by measuring their electrical transport characteristics with a source-measure unit (Keithley 2450) in dark and upon illumination. We use high-power fiber-coupled LED sources (Thorlabs) with 18 different wavelengths to study the spectral response. A spot of 67 $mm^2$ in area with a power of 30 mW is used for all the photocurrent measurements at different wavelengths. Figure 4a shows the current *vs.* voltage curve (*IV* hereafter) acquired for the device in the dark state and upon illumination with selected wavelengths showing a clear photoresponse. To verify if the $MoS_2$ channel is the main source of the photogenerated current we show in the Supporting Information the measurement on a graphite-on-paper device with a poor response to illumination. Figure 4b shows the current flowing through the device as a function of time when the illumination is switched ON and OFF to determine the response speed of the devices. When the illumination is switched ON, the photodetector shows an initial sharp response (faster than 0.2 s) followed by a slower response (~20-30 s) which indicates a superposition of different photocurrent mechanisms. The sharp response is typically observed in devices whose photocurrent generation is dominated by the photoconductive effect. Other photogeneration mechanisms like photogating or bolometric could be the source of the observed slow response.[31–35] The photocurrent can be determined by subtracting the dark current to the current under illumination. The responsivity, a common figure-of-merit that allows the comparison between different photodetectors, can be calculated as:

$$R = \frac{I_{ph}}{P} \cdot \frac{A_{spot}}{A_{sample}},$$

where $I_{ph}$ is the photocurrent, $P$ is the effective power, $A_{spot}$ is the area of the focused spot and $A_{spot}$ is the illuminated active area of the device. Figure 4c summarizes the responsivity of the device at different illumination wavelength in the 365 nm to 940 nm range. Interestingly, the $MoS_2$-on-paper photodetector shows a very broad spectral response with a shallow peak at 550-750 nm that matches the spectral range where the optical absorption of the $MoS_2$ film is enhanced due to the presence of the A and B excitonic resonances (see Figure 3b).[36,37] Although there are some reports about photoresponse in multilayer $MoS_2$ detectors working in the NIR, the typical response drops abruptly beyond 670 nm.[38,39] This spectral response beyond 670 nm also suggest the presence of other photogeneration mechanism like bolometric effect.

The responsivity reaches a value of ~1.5 μA/W, much smaller than that of $MoS_2$ nanodevices based on single crystals.[33,40,41] This reduced performance is expected for a macroscopic device formed by the overlap of small crystallites as a higher density of recombination centers is expected. Nonetheless, the responsivity value is comparable to that of devices based on liquid phase exfoliation or inkjet printing.[42,43] Note, that by operating the photodetector at higher temperatures the responsivity can be increased up to 10 μA/W (see the Supporting Information).

In order to get a deeper insight about the physical mechanism behind the photocurrent generation we studied the power and bias voltage dependence of the photocurrent. Figure 5a shows the current flowing in the devices as a function of time while the illumination is switched ON and OFF for different incident powers (going from 5 mW to 75 mW). Figure 5b displays the generated photocurrent as a function of the power density showing a marked linear relationship. The responsivity is thus almost independent on the power density (Figure 5c). Figure 5d and 5e show similar measurements to 5a and 5b respectively when varying the bias voltage instead of the power density. The photocurrent is linearly proportional to the bias voltage with increasingly higher responsivity for higher voltages. (Figure 5f). The linear power and bias dependence of the photocurrent points to a major contribution of the photoconductive effect, which would explain the initial sharp response of the devices to modulated light. The linear power dependence also rules out photogating as the source of the slow response component of the photogenerated current. In fact, photogating is characterized by a sublinear power dependence.[31–35] The bolometric effect, on the other hand, would be compatible with a linear power and bias dependence of the photocurrent and thus it could be the origin of the slow response component. Moreover, the bolometric effect would explain the broadband spectral response. In this scenario, the graphite electrodes would absorb light in a broad spectral range (even beyond that of $MoS_2$) increasing the temperature of the device, thus changing its resistance. In support to this scenario, in the Supporting Information we show that the $MoS_2$ device has a strong temperature dependent resistance.

**Paper-based single-pixel camera imaging device**

To further demonstrate the potential of these photodetectors, we mount them in a homebuilt single-pixel camera setup to image objects. The linear response to light of the paper-based photodetectors makes them ideal active elements in imaging. Figure 6a is a sketch of the experimental setup. Briefly, the object to be imaged is mounted on a motorized XY stage (Standa, 2x 8MT167S-25LS stages with 8SMC5-USB-B9-2 controller). A reflection-probe fiber bundle (RP29, Thorlabs) is used to illuminate and to probe the light reflected by the object under study (sample). This reflection-probe fiber bundle is bifurcated with three legs: one leg pointing towards the sample (sample leg), one leg carries light from a source towards the sample (light source leg) and another leg carries the light reflected by the sample to the photodetector (photodetector leg). By raster-scanning the object in the X and Y directions, one constructs a map of photocurrent that is linearly proportional to the reflectivity change in the object. Figure 6b compares the picture of a paper smiley acquired with a cell phone camera and a photocurrent map acquired with the $MoS_2$-on-paper photodetector single-pixel camera.

## Conclusions

In summary, we have demonstrated how van der Waals materials other than graphite can be used to draw devices on paper. In particular, we show how the layered structure of $MoS_2$ allows to deposit interconnected platelet traces on the surface of common paper by simply rubbing a $MoS_2$ crystal against a it. We also show how this simple method can be used to fabricate $MoS_2$-on-paper photodetectors with a remarkable broad spectral range. We have characterized the performance of these photodetectors finding that the photocurrent is generated by a superposition of photoconductive and bolometric effects, with responsivity values in the order of 1-2 µA/W (and even up to 10 µA/W when the device is operated at 70 ºC). Finally, we successfully demonstrate the potential use of these paper-based photodetectors by integrating one on a single-pixel camera setup to acquire images of objects.

## Conflicts of interest

There are no conflicts to declare.

## Acknowledgements

We would like to thank the Referees for their constructive input during the peer-review process, the final version of the manuscript has improved substantially thank to their suggestions. This project has received funding from the European Research Council (ERC) under the European Union's Horizon 2020 research and innovation programme (grant agreement number 755655, ERC-StG 2017 project 2D-TOPSENSE) and the European Union's Horizon 2020 research and innovation program under the Graphene Flagship (grant agreement number 785219, GrapheneCore2 project and grant agreement number 881603, GrapheneCore3 project). R.F. acknowledges the support from the Spanish Ministry of Economy, Industry and Competitiveness through a Juan de la Cierva-formación fellowship 2017 FJCI-2017-32919. We acknowledge support of the publication fee by the CSIC Open Access Publication Support Initiative through its Unit of Information Resources for Research (URICI).

## Notes and references

‡ During the elaboration of this manuscript we became aware of the pre-print by Nutting *et al*. [44] where the authors report the fabrication of paper-supported electronic devices based on different layered materials by means of abrasion of fine powder of the layered material against paper.

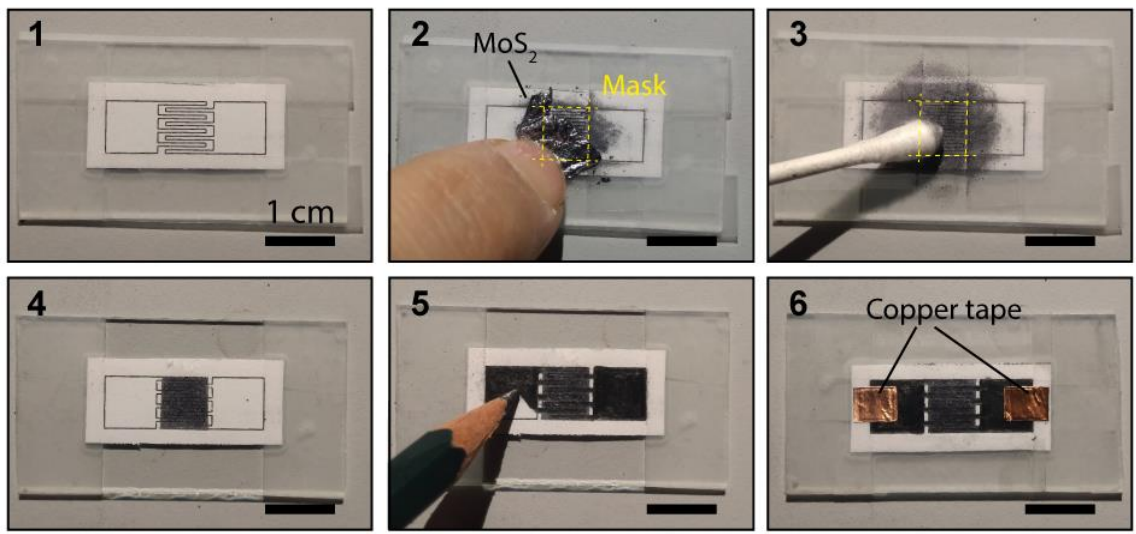

**Figure 1. Pictures of the fabrication process of a MoS$_2$-on-paper photodetector.** (1) The outline of the interdigitated electrodes is printed out in an office laser printer, the paper is cut and fixed onto a glass slide with adhesive tape. (2) A square mask is made in the device active area with adhesive tape and a MoS$_2$ crystal is rubbed against the bare paper area. (3) The drawn-MoS$_2$ is blurred with a cotton swab to improve the homogeneity. (4) After repeating the rubbing + blurring steps 4 times the mask is removed yielding to a very homogeneous MoS$_2$ square film. (5) The electrodes are drawn, following the printed outline, with a 4B pencil. (6) Two squares of copper tape are adhered to the graphite pads to allow soldering wires.

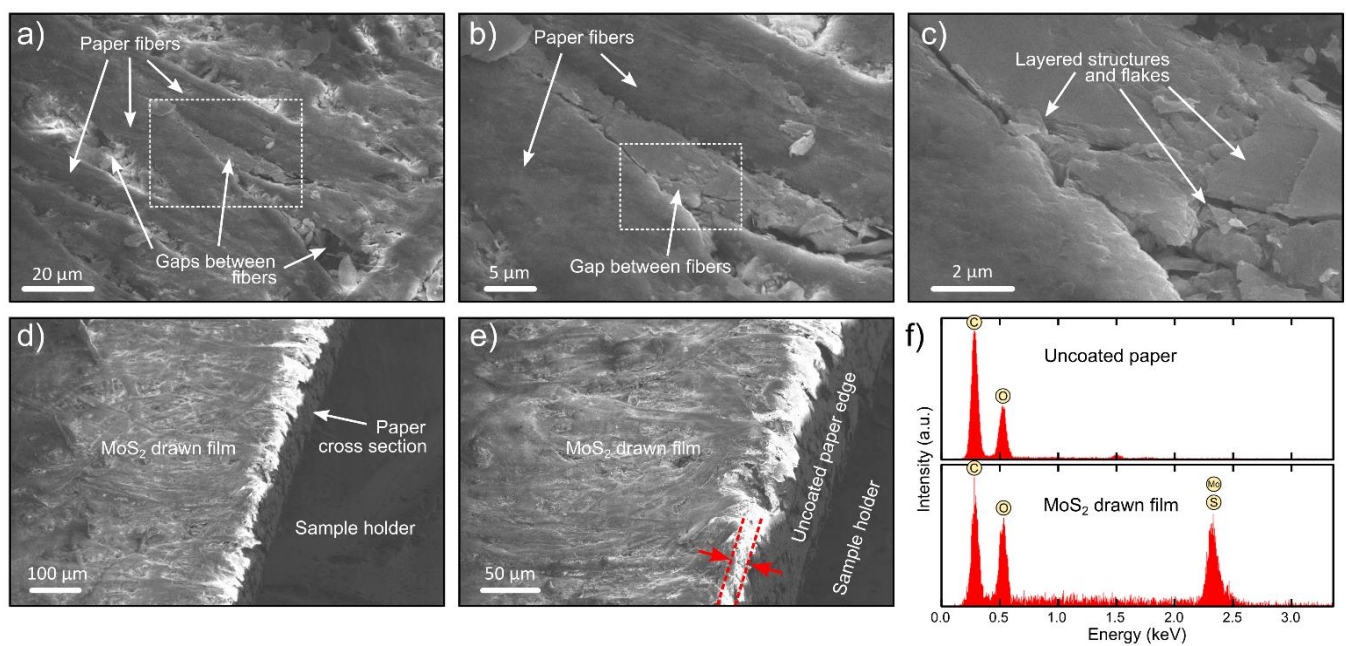

**Figure 2. Scanning electron microscopy (SEM) and energy dispersive X-ray (EDX) spectroscopy analysis of an as-drawn MoS$_2$ film on paper.** (a) to (c) SEM images with increasingly higher magnification showing the morphology of the MoS$_2$ film on paper. (c) Zoomed-in image of the layered structures and flakes that fill-in the gaps between the paper fibers. (d) and (e) High-angle SEM images of the sample cross-section displaying the uncoated paper edge and the MoS$_2$ film drawn on top. (e) SEM picture used for the estimation of the MoS$_2$ film thickness. (f) EDX analysis of the chemical composition of the MoS$_2$ film and the bare paper (for comparison).

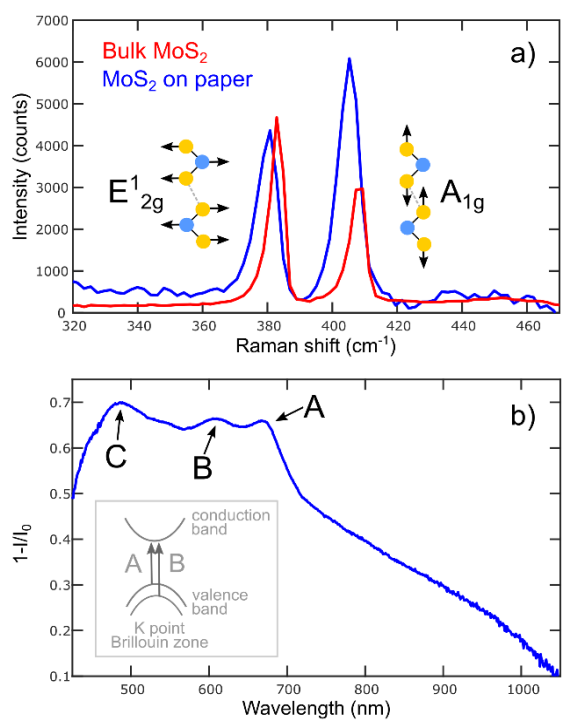

**Figure 3. Raman and transmittance spectroscopy characterization of an as-drawn MoS$_2$ film.** (a) Comparison between the Raman spectra acquired on a bulk MoS$_2$ crystal and that measured on a MoS$_2$-on-paper film (same film as in Fig. 2). (b) Transmittance spectrum (plotted as 1-transmittance) of the MoS$_2$-on-paper film. The A, B and C excitonic peaks have been highlighted. The inset in (b) shows a cartoon of the direct band transitions at the K point of the Brillouin zone that originate the A and B excitonic features.

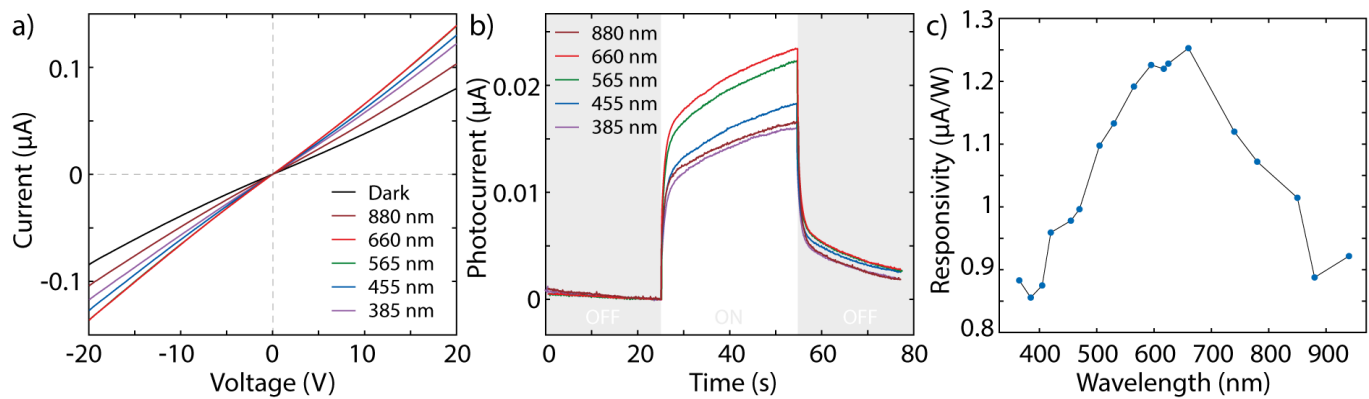

**Figure 4. Optoelectronic response of the as-drawn MoS$_2$ photodetectors.** (a) Current *vs.* voltage characteristics of the MoS$_2$-on-paper device in dark and upon illumination with selected illumination wavelengths (incident power 30 mW). (b) Photocurrent (current minus dark current value) flowing across the device (at a fixed bias voltage) as a function of time while the illumination with selected wavelengths is switched ON and OFF. (c) Responsivity spectrum of the device in the visible and near-infrared. Note: (b) and (c) measurements are carried out at $V_{bias}$ = 20 V and incident power of 30 mW.

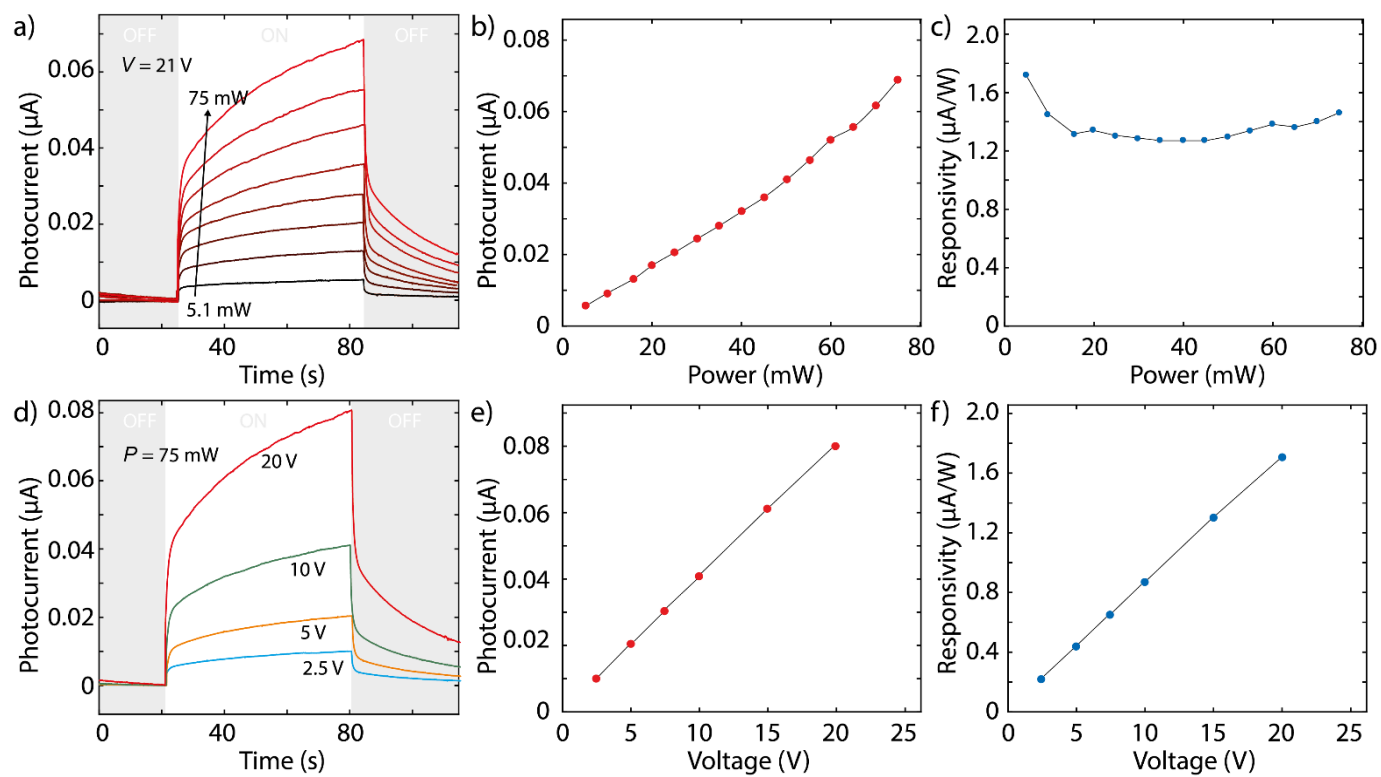

**Figure 5. Power and bias dependence of the photocurrent generation in the as-drawn MoS$_2$ photodetectors.** (a) and (b) Power dependence of the photocurrent (at a fixed bias voltage $V$ = 21 V). (c) Responsivity as a function of the incident power showing a rather constant value around 1.4 µA/W. (d) and (e) Bias voltage dependence of the generated photocurrent. (f) Bias voltage dependence of the responsivity of the device (at a fixed wavelength of 660 nm and fixed power of 75 mW).

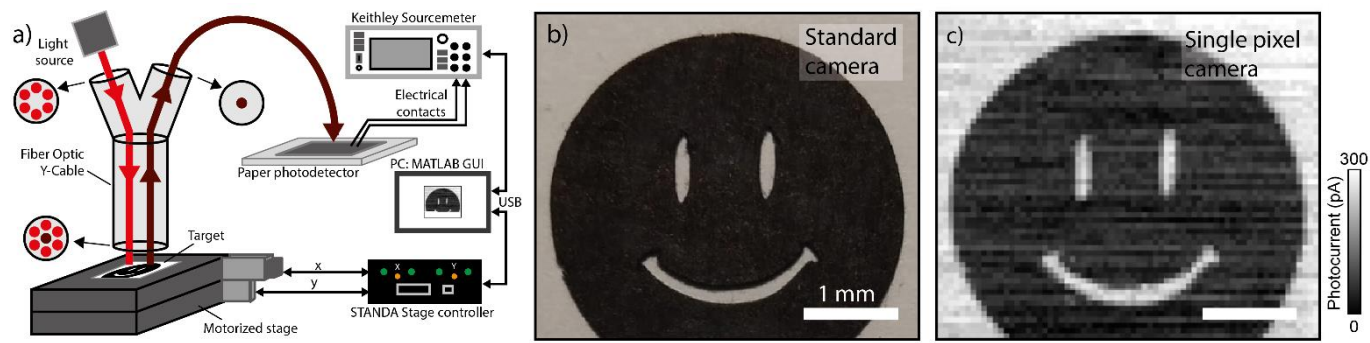

**Figure 6. Integration of the MoS$_2$-on-paper photodetector in a single-pixel camera system to acquire images.** (a) Schematic diagram of the experimental setup. (b) Image of the object under study acquired with a standard camera. (c) Image of the same object acquired with the single-pixel camera system based on a MoS$_2$-on-paper photodetector.

## Supporting Information:

## MoS$_2$-on-paper optoelectronics: drawing photodetectors with van der Waals semiconductors beyond graphite

*Ali Mazaheri[1,2], Martin Lee[3], Herre S.J. van der Zant[3], Riccardo Frisenda[1], Andres Castellanos-Gomez[1*]*

[1]Materials Science Factory. Instituto de Ciencia de Materiales de Madrid (ICMM-CSIC), Madrid, E-28049, Spain.

[2]Nanophysics research Lab., Department of Physics. University of Tehran, Tehran 14395, Iran.

[3]Kavli Institute of Nanoscience, Delft University of Technology, Lorentzweg 1, 2628 CJ, Delft, The Netherlands.

[*]andres.castellanos@csic.es

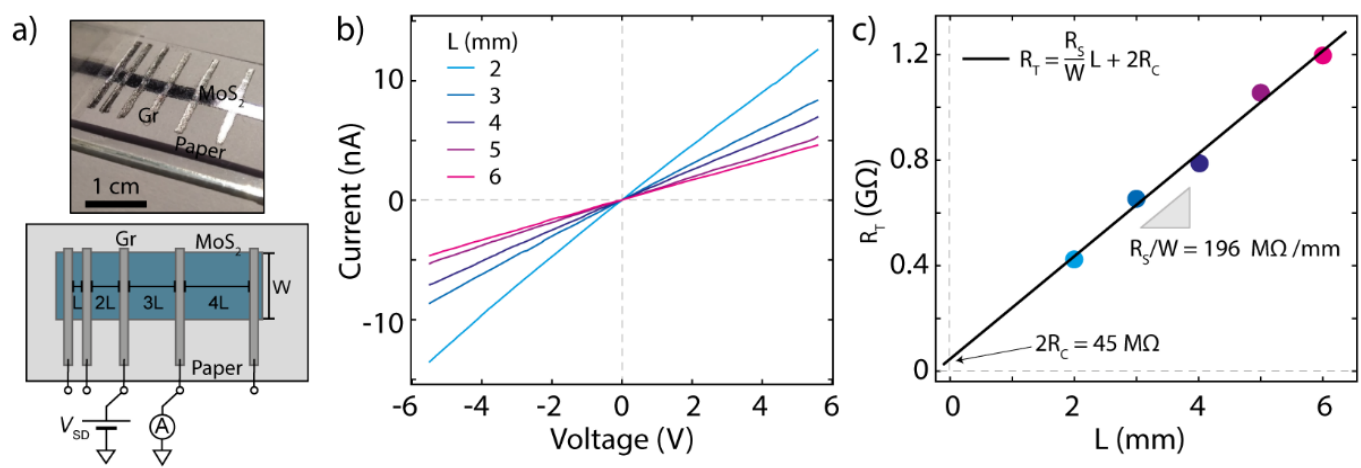

**Figure S1. Contact resistance measurement though the transfer length method.** (a) Picture and scheme of the fabricated device. The width (W) of the MoS$_2$ strip is 2 mm. (b) *IV* characteristics measured for different channel lengths. (c) Resistance *vs.* channel length. The experimental data follows a linear trend and the contact resistance can be found from the crossing of the linear trend with the vertical axis.

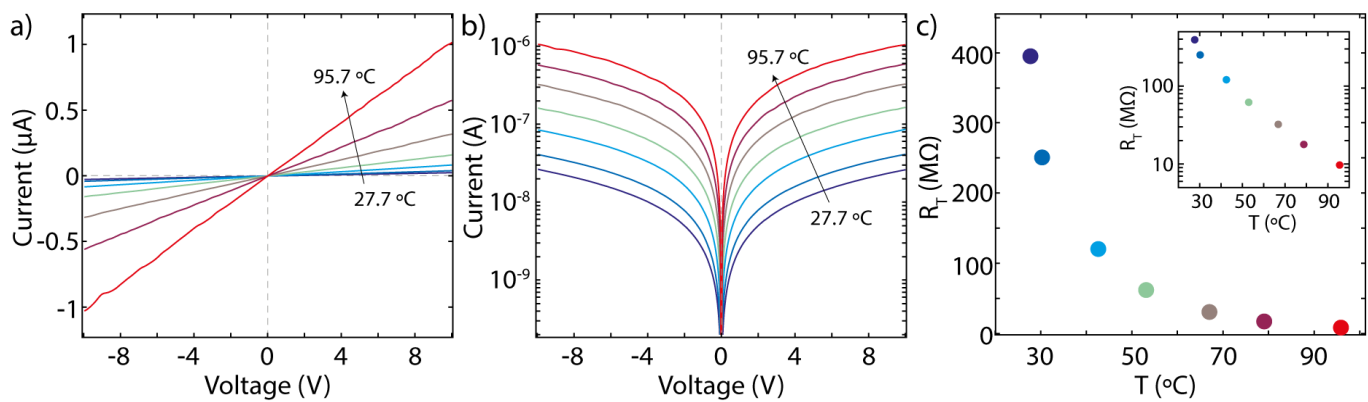

**Figure S2. Temperature dependent electrical characteristics.** (a) *IV* characteristics of a MoS$_2$ photodetector device acquired at different temperatures. (b) Absolute value of the current (in logarithmic scale) *vs.* bias voltage at different temperatures. This representation allows to better resolve the electrical characteristics change upon heating. (c) Resistance as a function of the temperature. (Inset in c) semilogarithmic plot of the temperature dependence of the resistance where the exponential decay of the resistance with temperature is evident.

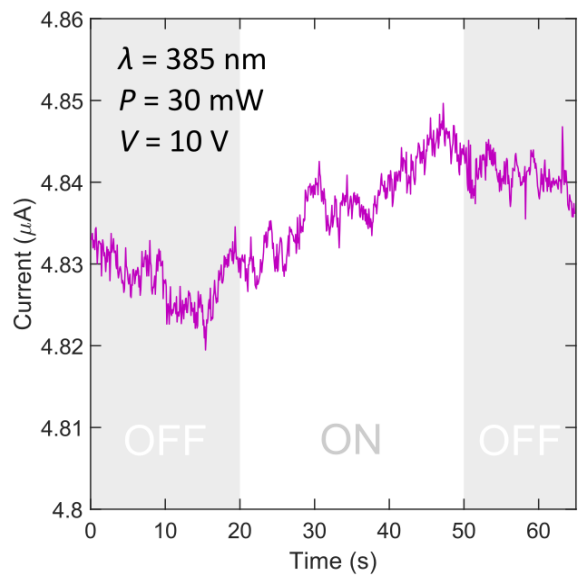

**Figure S3.** Current flowing across a graphite-on-paper device (similar to that shown in Figure 1 but with graphite channel instead of MoS$_2$), at a fixed bias voltage $V = 10V$, as a function of time while a 385 nm illumination source is switched ON and OFF. The graphite channel has been fabricated in a similar manner to the MoS$_2$ device.

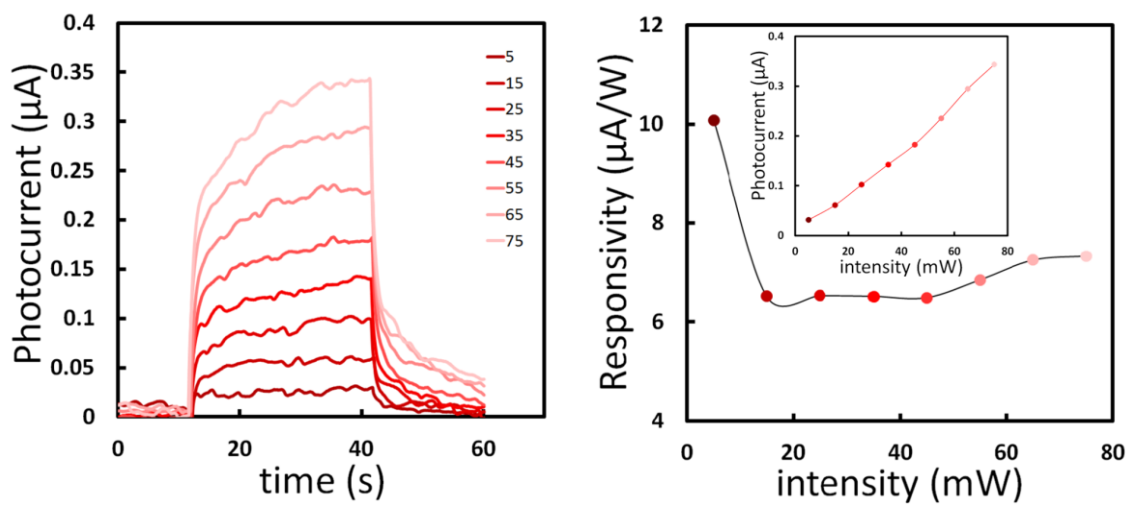

**Figure S4. Photoresponse measured at 70 ºC.** (left) Power dependence of the photocurrent (at a fixed bias voltage $V = 20$ V, wavelength $\lambda = 660$ nm and temperature $T = 70$ ºC). Illumination power ranges from 5 mw to 75 mW. (right) Responsivity as a function of the incident power showing a rather constant value around 7 µA/W. (inset) Photocurrent *vs.* incident power relationship.